\documentclass[letterpaper,british,english,reprint, pral]{revtex4-1}
\usepackage[T1]{fontenc}
\usepackage{textcomp}
\usepackage{amsmath}
\usepackage{amssymb}
\usepackage{graphicx}

\makeatletter

\pdfpageheight\paperheight
\pdfpagewidth\paperwidth

\makeatother

\usepackage{babel}
\begin{document}

\preprint{APS/123-QED}

\title{Alternative analysis of the delayed-choice quantum eraser with two entangled particles.}

\author{Sandro Faetti}
\email{sandro.faetti@unipi.it}

\selectlanguage{english}%

\affiliation{Department of Physics Enrico Fermi, Largo Pontecorvo 3, I-56127 Pisa,
Italy.}

\date{\today}
\begin{abstract}
According to Quantum Mechanics, the particles can exhibit  either
particle properties or wave properties depending on the experimental
set up (wave-particle dualism). A special behavior occurs for a system
of two entangled particles 1 and 2 that propagate along two different
directions in the space. In such a case, the entangled particle 2
should exhibit either the particle behavior or the wave behavior
depending on the kind of measurement that is performed on particle
1 whatever is the actual distance between the two particles. The apparently
surprising fact is that the ``choice'' of what kind of measurement
is  performed on particle 1 can be also made when particle 2 has
been already detected (delayed-choice quantum eraser). These theoretical
predictions have been confirmed by the experiments
and  seem to suggest that a future measurement can affect a past event.  Recently, both the concepts of  "delayed choice" and of "quantum erasure" have been criticized by Ellerman and by Kastner.  In this paper we  propose an alternative  analysis of the delayed-choice quantum 
eraser with two entangled particles and we show that it does not imply an inversion of the natural cause-effect temporal order of the physical events. 
\begin{description}
\item [{PACS~numbers}] 03.65.Ud, 03.65.-w, 03.65.Ta,42.50-p{\small \par}
\end{description}
\end{abstract}



\maketitle

\section*{Introduction: the quantum eraser.}

According to Quantum Mechanics, particles can exhibit either wave
properties or particle properties (wave-particle dualism) depending
on the experimental set up\citep{Bohr1928}. 
The wave-particle duality leads to many counterintuitive predictions
as well as, for instance the delayed-choice quantum erasure that has
been the object of many theoretical and experimental investigations
and of an interesting review paper\citep{RevModPhys.88.015005}. We remind
the reader to this reference for details and for an extended bibliography.
Recently Ellerman\citep{Ellerman} and Kastner\citep{Kastner}   have strongly criticized the
concepts of  ``delayed-choice'' and  ``erasure''.

\begin{figure}[h]
\centering{}\includegraphics[scale=0.60]{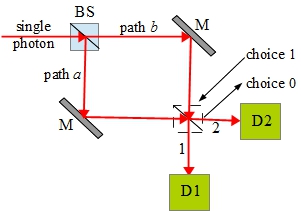}\caption{\label{fig.1} A single photon impinges on the symmetric 50\% beam
splitter BS, is reflected by mirrors M and is detected by the single
photon detectors D1 ans D2. A second beam splitter (dashed in the
figure) can  be either inserted (choice "1") or removed (choice "0"). If the beam splitter is removed, the clicking of either detector D1 or detector D2 allow
to know which path ($a$ or $b$) has been followed by the photon. In the other case   the which-path information is erased and
 interference occurs.}
\end{figure}
Here we report an  alternative analysis  of the delayed choice quantum eraser with two entangled particles and we
show that it does not imply the counterintuitive
inversion of the natural temporal order of the physical events. A famous delayed-choice gedanken experiment was proposed by Wheeler
in 1984\citep{Wheeler} and is shown in Figure \ref{fig.1}. A single photon impinges on the first beam splitter of a
Mach-Zender interferometer.  If the second beam splitter is removed
(choice ``0''), the detection of the photon by either detector D1
or D2 allows to know which is the path followed by the photon (which-path information and particle behavior).
In such a case, no interference occurs and the clicks of detectors
D1 and D2 are completely indipendent on the optical path difference
between the two arms of the interferometer. The insertion of the second
beam splitter (choice ``1'') erases the which-path information and
interference occurs at outputs 1 and 2 (wave behavior). In such a
case, the clicks of detectors D1 and D2 depend on the difference of
optical paths between the two arms of the interferometer. Wheeler
emphasized an  important feature: if paths $a$ and $b$ are sufficiently
long,  the choice of either inserting a second beam splitter or
not can be made after the photon has already passed through the first beam
splitter. Then, by such a delayed-choice experiment, we get the counterintuitive
result that the future choice seems to affect the past history of
the photon. In particular, if the beam splitter is inserted, the which-path information is erased.  However, Wheeler himself wrote: \textquotedblleft{} in
actuality it is wrong to talk of the route of a photon and it makes
no sense to talk of the phenomenon until it has been brought to a
close and irreversible act of amplification\textquotedblright . Other
delayed-choice gedanken experiments have been proposed in the literature
and the Quantum Mechanics predictions have been always verified in
successive real experiments \citep{RevModPhys.88.015005}. 

\begin{figure*}[t]
\begin{centering}
\includegraphics[scale=0.60]{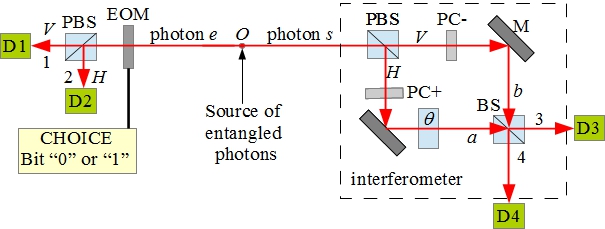}
\par\end{centering}
\caption{\label{Fig 2}\foreignlanguage{british}{ }Scheme of a delayed-choice
quantum eraser with two entangled photons (the environment photon $e$ and the system photon $s$). PBS are polarizing beam
splitters, M are mirrors, BS is a symmetric 50\% beam splitter. PC+
and PC- are polarization controllers that rotate the vertical and
horizontal polarizations \emph{V} and \emph{H} by 45\textdegree{}
in opposite directions to align them
along the same diagonal axis to make  possible interference. The
dashed rectangle delimits a modified Mach-Zehnder interferometer where
the usal first beam splitter is replaced by a polarizing beam splitter.
$\theta$ denotes a dephasing plate that introduces the phase delay
$\theta$ between the arms $a$ and $b$  of the interferometer. EOM
is an electro-optic modulator. Choice ``0'' correspond to the case
where the EOM is ``off'', while choice ``1'' corresponds to the
case where the EOM is ``on''.}
\end{figure*}

Scully et al.\citep{Scully1,Scully2} investigated an  atom-photon entangled
state and proposed the so-called delayed-choice quantum eraser\citep{Kim} that is the object of  our  successive analysis. After
this pioneering work, other configurations of entangled systems have
been investigated (see\citep{RevModPhys.88.015005}) and, in particular
photon-photon entangled configurations. Figure 2 shows the case of
two polarization entangled photons, the environment (or idler) photon
$e$ and the system (or signal) photon $s$, that are generated at
point \emph{O} by spontaneous parametric down conversion \citep{Kwiat} and
propagate along different directions in the space. This kind of apparatus has been used, for instance,  in reference\citep{erasure}.
 Photons $e$ and $s$ are in
the polarization maximally entangled state 

\begin{equation}
|\psi\text{\textrangle}=\frac{1}{\sqrt{2}}\left(|H\text{\textrangle}_{e}|V\text{\textrangle}_{s}+|V\text{\textrangle}_{e}|H\text{\textrangle}_{s}\right),\label{eq:Smax-3}
\end{equation}
where \emph{H} and \emph{V} denote horizontal and vertical polarization,
respectively. The entangled state represents a very special state
predicted by Quantum Mechanics that has no equivalent in classical and semi-classical physics. In
particular, each entangled photon  has  not a definite polarization
(linear, circular or elliptical polarization) but the   polarizations of the two entangled photons 
are strictly correlated (EPR correlations). For instance,
if photon $e$ is found in the vertically polarized state $|V\text{\textrangle}_{e}$,
photon $s$ is found in the horizontally polarized state $|H\text{\textrangle}_{s}$.
When the system photon $s$ passes through the polarizing beam splitter PBS
of the interferometer and gets the inputs of the successive beam splitter BS, the polarization entangled state becomes the
polarization-path hybrid state

\begin{equation}
|\psi\text{\textrangle}=\frac{1}{\sqrt{2}}\left(|H\text{\textrangle}_{e}|b\text{\textrangle}_{s}+i\,e^{i\theta}|V\text{\textrangle}_{e}|a\rangle\rangle_{s}\right),\label{eq:Smax-3-1}
\end{equation}
where we use the reduced forms $|a\text{\textrangle}_{s}$ and $|b\text{\textrangle}_{s}$
to denote a one-photon state at the $a$-input of the
beam splitter BS  with a vacuum state at the $b$-input
($|a\text{\textrangle}_{s}=|1\text{\textrangle}_{a}|0\text{\textrangle}_{b}$)
and a one-photon state at the $b$-input with a vacuum state at  the
$a$-input ($|b\text{\textrangle}_{s}=|0\text{\textrangle}_{a}|1\text{\textrangle}_{b}$),
respectively. The phase contributions $i$ and $e^{i\theta}$ appearing
in Eq.(\ref{eq:Smax-3-1}) are due to the PBS transfer operator (see Eq.(1.21)
in \citep{BEAMSPLITTER} with $t_{H}=0,t_{V}=1,r_{H}=1$ and $r_{V}=0$) \footnote {A coefficient $ i$   coming from the contibution $i r_{H}$ was lost in the successive  Eq.(1.22) of  \citep{BEAMSPLITTER} }
and to the dephasing plate operator, respectively. According to
Eq.(\ref{eq:Smax-3-1}), the  polarizations \emph{H} and \emph{V}
of the environment photon are strictly related to the paths $a$ and
$b$ followed by the system photon in the interferometer. Then, it
is usually concluded that  the environment photon $e$ carries the which-path information on
the system photon $s$ although this conclusion has been strongly criticized  by R.E.Kastner\citep{Kastner}.
Choice \textquotedblleft 0\textquotedblright{} corresponds to the
case where the electro-optic modulator EOM is \textquotedblleft off\textquotedblright{}
and the polarization of the environment photon $e$ is not modified
by it. Then, the PBS in front to the EOM in figure \ref{Fig 2} sends
the vertically  and horizontally polarized photons \emph{V} and \emph{H} to
outputs 1 and 2, respectively. In this case the 
components \emph{V} and \emph{H}  of the polarization of the environment photon are measured
by detectors D1 and D2, respectively. Choice \textquotedblleft 1\textquotedblright{}
corresponds to the case where the electro-optic modulator is \textquotedblleft on\textquotedblright{}
and its dephasing is set in such a way that right-hand ($R$)
and left-hand ($L$) circularly  polarized photons impinging on the electo-optic modulator EOM are changed by it
to linear vertically ($V$) and horizontally ($H$) polarized photons, respectively.
In such a case the optical system (EOM + PBS) sends the incident right
and left hand circular polarizations $R$ and $L$ to outputs 1 and
2 of the polarizing beam splitter, respectively.  Then,  the system EOM+PBS behaves as a beam splitter for the circular polarizations in this case.  In the case of choice
\textquotedblleft 0\textquotedblright , if the environment photon
is found to be in the state $|V\text{\textrangle}_{e}$ (or $|H\text{\textrangle}_{e}$), the system photon must be in the state $|a\text{\textrangle}_{s}$
(or $|b\text{\textrangle}_{s}$) and, thus, the current  interpretation is that  the photon has  passed
only in  path $a$ (or $b$). In this case,  no interference can occur and  the coincidences between detectors Di ($i$=1,2) and Dj ($j$=3,4) are independent of the dephasing $\theta$  between the two arms of the interferometer. A completely
different behavior is predicted  if the right-hand circular polarization
($R$) and the left-hand circular polarization ($L$) of the environment
photon are measured (choice \textquotedblleft 1\textquotedblright ).
Rewriting the entangled state of Eq.(\ref{eq:Smax-3-1}) in the new
circular orthonormal basis $|R\text{\textrangle}_{e}=\frac{1}{\sqrt{2}}\left(|H\text{\textrangle}_{e}+i\,|V\text{\textrangle}_{e}\right)$
and $|L\text{\textrangle}_{e}=\frac{1}{\sqrt{2}}\left(|H\text{\textrangle}_{e}-i\,|V\text{\textrangle}_{e}\right)$
and disregarding a common multiplicative coefficient \emph{i}, one
gets the alternative expression

\begin{eqnarray}
|\psi\rangle = &&\frac{1}{2}\left[|L\rangle_{e}\left(-e^{i\theta}|a\rangle_{s}+|b\rangle_{s}\right) \right. \nonumber \\
 && \left. +|R\rangle_{e|}\left(e^{i\theta}|a>_{s}+|b>_{s}\right)\right].\label{eq:Smax-3-1-1}
\end{eqnarray}
From Eq.(\ref{eq:Smax-3-1-1}) we infer that, if the environment photon is found in the state either $|R\text{\textrangle}_{e}$
or $|L\text{\textrangle}_{e}$, the which-path information is lost
and interference occurs. Then, the choice  of measuring  the  circular polarizations \emph{R} and
\emph{L} of the environment photon (choice ``1'') seems to produce the erasure of  the which-path information. At the final beam splitter BS, $|a\text{\textrangle}_{s}\rightarrow\frac{1}{\sqrt{2}}\left(|3\text{\textrangle}_{s}+i\,|4\text{\textrangle}_{s}\right)$
and $|b\text{\textrangle}_{s}\rightarrow\frac{1}{\sqrt{2}}\left(|4\text{\textrangle}_{s}+i\,|3\text{\textrangle}_{s}\right)$
where $|3\text{\textrangle}_{s}$ and $|4\text{\textrangle}_{s}$
shortly denote one-photon states at outputs 3 and 4, respectively. Then, the
entangled state in Eq.(\ref{eq:Smax-3-1-1}) becomes the hybrid entangled state

\begin{eqnarray}
|\psi'\text{\textrangle}=&&\frac{1}{\sqrt{2}}\left[i\,\cos\alpha|L\text{\textrangle}_{e}|3\text{\textrangle}_{s}+\sin\alpha|R\text{\textrangle}_{e}|3\text{\textrangle}_{s} \right.\nonumber \\
 &&\left. -i\,\sin\alpha|L\text{\textrangle}_{e}|4\text{\textrangle}_{s}+\cos\alpha|R\text{\textrangle}_{e}|4\text{\textrangle}_{s}\right],\label{eq:Smax-3-1-1-1}
\end{eqnarray}
where $\alpha$ is the phase coefficient

\begin{equation}
\alpha=\frac{\theta}{2}+\frac{\pi}{4}.\label{eq:Smax-3-1-1-2}
\end{equation}
A common inessential multiplicative phase factor $e^{i\alpha}$ was
disregarded in Eq.(\ref{eq:Smax-3-1-1-1}). We remind here that, for
choice "1", detectors D1 and D2 collect the right hand circularly polarized
photons and the left hand circularly polarized photons, respectively. Then, from Eq.(\ref{eq:Smax-3-1-1-1}) we infer that the probabilities
$p_{ij}$($i$ = 1,2 and $j$ =3,4) of having correlated clicks of
detectors Di and Dj ($i$ = 1,2 and $j$ =3,4) are

\begin{eqnarray}
p_{13}&=& \frac{1}{2}\sin^{2}\alpha,\label{eq:Smax-3-1-1-2-1}\\
p_{23}&=&\frac{1}{2}\cos^{2}\alpha,\label{eq:Smax-3-1-1-2-2}\\
p_{14}&=&\frac{1}{2}\cos^{2}\alpha,\label{eq:Smax-3-1-1-2-3}\\
p_{24}&=&\frac{1}{2}\sin^{2}\alpha.\label{eq:Smax-3-1-1-2-4}
\end{eqnarray}
The dependence of these probabilities on phase $\alpha$  evidences
the wave behavior of the photons. It is important to notice that the
probabilities that the system photon is detected by either detector
D3 or detector D4 are $p_{3}=p_{13}+p_{23}=\frac{1}{2}$ and $p_{4}=p_{14}+p_{24}=\frac{1}{2}$
that are independent of phase $\alpha$ as in the case of choice "0". Then, the interference contributions
in Eqs.(\ref{eq:Smax-3-1-1-2-1})-(\ref{eq:Smax-3-1-1-2-4}) can be
only observed if correlated measurements between detectors D1 (or
D2) and D3 (or D4) are performed for each couple of entangled photons.
The delayed-choice quantum erasure  occurs if the paths
of the environment photon $e$ to reach the polarizing elements (EOM
and PBS) and the successive detectors D1 and D2 are much longer than
the paths needed to the system photon $s$ to reach the interferometer
and detectors D3 and D4. In these conditions, the \textquotedblleft choice\textquotedblright{}
on what kind of polarization of the environment photon is measured
can be made when the system photon has already passed through the
interferometer and has been already detected (and registered) by either
detector D3 or detector D4. Then, the analysis above leads
to the counterintuitive conclusion that the successive choice decides
if the system photon behaves as a particle or as a wave after it  has been already detected and registered.  We note
here that the interpretation above suffers of an implicit realism
hypothesis while, according to Bohr, \textquotedblleft{} in Quantum
Mechanics no elementary phenomenon is a phenomenon until it is a registered
(observed) phenomenon\textquotedblright . According to the above sentence, the interpretation of the experimental results in terms of erasure of the which-path information  can be object of some criticism.  However, the experiments  with entangled particles
seem to suggest that an inversion of the causal temporal order also  occurs as far as the  correlations between the two physical events ( the clicks of the detectors) are concerned.  This latter behavior appears  to be very counterintuitive.
 Reference\citep{RevModPhys.88.015005} strongly stressed that the choice must be totally random  in an
ideal experiment and the passage
of the system photon through the first beam splitter and the \textquotedblleft choice\textquotedblright{}
must be space-like events. These requirements avoid possible causal
interpretations of the observed phenomena in terms of subluminal or
luminal communications between the events. Experiments satisfying
these conditions have been carried out successfully by Xiao-Song Ma
et al.\citep{erasure} and by F. Kaiser et al.\citep{Kaiser}.

The analysis reported in this introductive  Section  follows  that already  given   in reference\citep{erasure}.    In Section I we reanalyze the delayed-choice quantum eraser  with two entengled photons using
a different point of view and we show that the delayed-choice quantum
erasure admits an alternative interpretation that does not lead to the
usual counterintuitive conclusion that a future choice affects the
result of a past measurement in agreement with the conclusions of
Elleman\citep{Ellerman} and Kastner\citep{Kastner}. We show that the delayed-choice
quantum erasure it strictly related to the non-local character of
the Quantum Mechanics that has been evidenced by the EPR paradox\citep{EPR}
(Einstein, Podolsky and Rosen) and confirmed by
many successive EPR experiments (see, for instance, the most recent loophole-free experiments\citep{Loophole1,Loophole2,Loophole4}). 

\section{The alternative analysis of the delayed-choice quantum eraser.}

In this Section we analyzes the delayed-choice eraser using an alternative
approach. For the successive analysis it is convenient to rewrite
the entangled state in eq.(\ref{eq:Smax-3}) in terms of the orthonormal
basis of elliptically polarized states 

\begin{eqnarray}
|E\text{\textrangle}_{e,s}=\frac{1}{\sqrt{2}}\left[|H\text{\textrangle}_{e,s}+e^{i\theta}|V\text{\textrangle}_{e,s}\right],\label{eq:Smax-3-1-1-3}\\
|E^{\bot}\text{\textrangle}_{e,s}=\frac{1}{\sqrt{2}}\left[|H\text{\textrangle}_{e,s}-e^{i\theta}|V\text{\textrangle}_{e,s}\right],\label{eq:Smax-3-1-1-3-1}
\end{eqnarray}
where subscripts $e$ and $s$ refer to the environment photon and
the system photon, respectively and $\theta$ is the delay contribution
on the $a$-arm of the interferometer in figure 2. The expression
of the entangled state in the new elliptical basis is (up to an inessential
multiplicative phase factor $e^{-i\theta}$):

\begin{equation}
|\psi\text{\textrangle}=\frac{1}{\sqrt{2}}\left[|E\text{\textrangle}_{e}|E\text{\textrangle}_{s}-|E^{\bot}\text{\textrangle}_{e}|E^{\bot}\text{\textrangle}_{s}\right].\label{eq:Smax-3-1-1-3-2}
\end{equation}
Our successive analysis is strictly related to an important property
of the modified Mach Zehnder interferometer of figure \ref{Fig 2}:
this interferometer behaves as a elliptical polarizing beam splitter. The
standard linear polarizing beam splitter sends an input photon that
is in the linearly polarized  state $|V\text{\textrangle}$ at one of the two outputs
of the polarizing beam splitter and an input photon that is in the
$|H\text{\textrangle}$ state at the other output. Analogously, using the procedures of  Quantum Optics,  it
can be easily shown that the interferometer of figure 2 sends the
elliptically polarized input state $|E\text{\textrangle}_{s}$ at output 3 of the final beam
splitter BS and the input state $|E^{\bot}\text{\textrangle}_{s}$
at output 4. In particular,  we get $|E\text{\textrangle}_{s}\longrightarrow i\,|3\text{\textrangle}_{s}$
and $|E^{\bot}\text{\textrangle}_{s}\longrightarrow-\,|4\text{\textrangle}_{s}$
at up an  inessential common multiplicative coefficient $e^{i\theta}$.
We emphasize  here that in both the cases (either incident $|E\text{\textrangle}_{s}$
or $|E^{\bot}\text{\textrangle}_{s}$) the system photon propagates
in both the arms of the interferometer leading to a fully constructive
interference at one of the two outputs (either 3 or 4) and a fully
destructive interference at the other output.  If a photon is detected by detector D3 ( or D4) it necessarily means that this photon has passed both the arms of the interferometer leading to a constructive interference at output 3 (or 4) of the interferometer (wave behavior). As soon as the system
photon $s$ exits from the interferometer, the incident polarization
entangled state in Eq.(\ref{eq:Smax-3-1-1-3-2}) becomes the hybrid entangled state:
\begin{equation}
|\psi_{C}>=\frac{1}{\sqrt{2}}\left[i |E\text{\textrangle}_{e}|3\text{\textrangle}_{s}+|E^{\bot}\text{\textrangle}_{e}|4{\textrangle}_{s}\right]].\label{eq:Smax-3-1-1-3-2-1}
\end{equation}
This means that if detector D3 (or D4) clicks, the state of the environment
photon collapses to the elliptical state $|E\text{\textrangle}_{e}$
(or $|E^{\bot}\text{\textrangle}_{e}$) and, thus, it is not surprising
that the correspondent successive clicks of detectors D1 and D2 will
depend on what kind of polarization of the environment photon  is measured (either linear or
circular). According to this alternative point of view, it is not
the delayed-choice that affects the past history of the system photon
but is the detection of the system photon by one of the detectors
D3 and D4 that determines the result of the successive polarization
measurement performed on the environment photon. No inversion of the temporal order occurs at all and the successive
\textquotedblleft choice\textquotedblright{} does not affect the
detection of the system photon. If the future choice is \textquotedblleft 0\textquotedblright ,
the  polarizations $V$ and $H$ of the environment  photon  $e$  are measured by detectors
D1 and D2, respectively. From Eq.(\ref{eq:Smax-3-1-1-3-2-1}) we infer that  the detection of the system
photon  $s$  by detector either D3 or D4  implies that the corresponding environment photon   has collapsed  to either the elliptically polarized state $|E\text{\textrangle}_{e}=\frac{1}{\sqrt{2}}\left[|H\text{\textrangle}_{e}+e^{i\theta}|V\text{\textrangle}_{e}\right]$
or $|E^{\bot}\text{\textrangle}_{e}=\frac{1}{\sqrt{2}}\left[|H\text{\textrangle}_{e}-e^{i\theta}|V\text{\textrangle}_{e}\right]$, respectively.
Substituting these latter expressions in Eq.(\ref{eq:Smax-3-1-1-3-2-1})
we get an alternative expression of  the photon state $|\psi_{C}\text{\textrangle}$ in terms of
the horizontally and vertically polarized states of the environment
photon:
\begin{eqnarray}
|\psi_{C}\text{\textrangle}=&&\frac{1}{2}\left[i\,|H\text{\textrangle}_{e}|3\text{\textrangle}_{s}+i\,e^{i\theta}|V\text{\textrangle}_{e}|3\text{\textrangle}_{s}\right. \nonumber \\
&&\left. +|H\text{\textrangle}_{e}|4\text{\textrangle}_{s}-e^{i\theta}|V\text{\textrangle}_{e}|4\text{\textrangle}_{s}\right].\label{eq:Smax-3-1-1-3-2-1-1}
\end{eqnarray}
We remind that, in the case of choice  \textquotedblleft 0\textquotedblright, the environment photons in the  states $|V\text{\textrangle}_{e}$
and $|H\text{\textrangle}_{e}$ are collected by detectors
D1 and D2, respectively while the system photons in the states $|3\text{\textrangle}_{s}$
and $|4\text{\textrangle}_{s}$ are collected by detectors D3 and D4,
respectively. Then, from Eq.(\ref{eq:Smax-3-1-1-3-2-1-1}) we infer that the probabilities of finding \textquotedblleft correlated\textquotedblright{}
clicks of detectors Di ($i$ = 1,2) and Dj ($j$ = 3,4) are given
by p$_{ij}=\frac{1}{4}$ and are independent of $\theta.$ This is
just the same result already obtained in the introduction but, now,
the physical interpretation of the phenomenon is not the same. In
the introduction, the detection of the  polarizations  $V$ and $H$
of the environment photon forced the system photon to pass through
only one arm of the interferometer (which-path information). This interpretation
could be  appropriate to describe the case where the first measurement
is performed on the envelopment photon althogh some remark is needed (see the note \footnote {Suppose, for instance, that the environment photon is found from the measurement to be in the horizontally polarized state $|H\text{\textrangle}_{e}$. According to Eq.(\ref{eq:Smax-3}), this measurement induces the collapse of the system photon to the  state $|V\text{\textrangle}_{s}$ and, consequently,  the photon passes only through the $ b$-arm of the interferometer. Then, we can infer that the measurement of  the $H$-polarization of the environment photon gives the which-path information on the system photon.  However it has to be noted that this  information is lost when the system photon is detected by one of  detectors D3 and D4 because it has been shown that  the detection by one of the two detectors at the outputs of the interferometer can occur only if the photon has passed through both the paths  $a$ and  $b$ (note that a photon propagating along one arm of the interferometer can be always considered in a suitable superposition of photon states propagating through both the arms and corresponding at the two incident elliptical polarized states    $|E\text{\textrangle}_{s}$
and $|E^{\bot}\text{\textrangle}_{s}$).Then, the detection of the photon by detector D3 (or D4) comes from a costructive interference at outputs 3 (or 4) and evidences the wave behavior of the photon. }). On the contrary, in the delayed-choice
case, the first detection of the system photon by one of detectors
D3 and D4 implies that the system photon has passed in both the interferometer
arms giving a total constructive interference at one of the two outputs
3 and 4, respectively. Then, the which-path information is not allowed in this case.  In both these cases (delayed choice and not delayed choice),
the same final probabilities of \textquotedblleft correlated\textquotedblright{}
clicks p$_{ij}=\frac{1}{4}$ are predicted to occur but the physical interpretation is different. If the choice
\textquotedblleft 1\textquotedblright{} is made, the circular $R$
and $L$ polarizations of the environment photon  $e$  are measured. The elliptical states $|E\text{\textrangle}_{e}$
and $|E^{\bot}\text{\textrangle}_{e}$ in terms of the circular basis
vectors $|R\text{\textrangle}_{e}=\frac{1}{\sqrt{2}}\left[|H\text{\textrangle}_{e}+i\,|V\text{\textrangle}_{e}\right]$
and $|L\text{\textrangle}_{e}=\frac{1}{\sqrt{2}}\left[|H\text{\textrangle}_{e}-i\,|V\text{\textrangle}_{e}\right]$
are (up to  a common inessential phase factor $e^{i\alpha}$ )

\begin{eqnarray}
|E\text{\textrangle}_{e}=\left[-i\,\sin\alpha|R\text{\textrangle}_{e}+\cos\alpha |L\text{\textrangle}_{e}\right],\label{eq:Smax-3-1-1-3-3}\\
|E^{\bot}\text{\textrangle}_{e}=\left[\cos\alpha|R\text{\textrangle}_{e}-i\,\sin\alpha|L\text{\textrangle}_{e}\right].\label{eq:Smax-3-1-1-3-1-1}
\end{eqnarray}
Substituting these expressions in Eq.(\ref{eq:Smax-3-1-1-3-2-1})
we get the state $|\psi_{C}\text{\textrangle}$ in terms of $|R\text{\textrangle}_{e}$
and $|L\text{\textrangle}_{e}$ :

\begin{eqnarray}
|\psi_{C}\text{\textrangle}= &&\frac{1}{\sqrt{2}}\left[i\,\cos\alpha|L\text{\textrangle}_{e}|3\text{\textrangle}_{s}+\sin\alpha|R\text{\textrangle}_{e}|3\text{\textrangle}_{s}\right.\nonumber \\
   &&\left. -i\,\sin\alpha|L\text{\textrangle}_{e}|4\text{\textrangle}_{s}+\cos\alpha|R\text{\textrangle}_{e}|4\text{\textrangle}_{s}\right].\label{eq:Smax-3-1-1-3-2-1-1-1}
\end{eqnarray}
The latter expression coincides with equation (\ref{eq:Smax-3-1-1-1})
already obtained in the Introduction and, thus, also in this case
the  probabilities p$_{ij}$ given in eqs.(\ref{eq:Smax-3-1-1-2-1})-(\ref{eq:Smax-3-1-1-2-4})
are immediately recovered. The main difference is that, now, it is
the first detection at detectors D3 and D4 that produces the collapse
of the entangled state and the successive correlated detections of
detectors D1 and D2. Although the calculations made in the introduction
and in Section I lead to the same final results for the probabilities $p_{ij}$, we  think that the approach of Section I has
to be physically preferred in the case of a delayed-choice experiment with entangled particles. In fact,
according to Eq.(\ref{eq:Smax-3-1-1-3-2-1-1}), it is the detection of the system photon by either detector
D3 or D4 that induces the  collapse of the environment photon  to either $|E\text{\textrangle}_{e}$
 or $|E^{\bot}\text{\textrangle}_{e}$, respectively.
For instance, according to Eq.(\ref{eq:Smax-3-1-1-3-2-1}), if detector
D3 clicks it means that the environment photon has collapsed to the
elliptical state $|E\text{\textrangle}_{e}$. This collapse of the
environment photon to the state $|E\text{\textrangle}_{e}$ represents
a well defined physical event  that occurs before the detection of the environment photon. In fact, the environment
photon will be always  found   in the polarizion state   $|E\text{\textrangle}_{e}$  if a successive measurement of
the elliptical polarizations  $E$  or $E^{\bot}$ is performed.
We emphazize  here that the interferometer of figure 2 is  equivalent
to an elliptical polarizing beam splitter  and, thus, the experiment shown in figure
2 is analogous to a typical  EPR experiment where the polarization
correlations between the entangled photons are measured \citep{Loophole1,Loophole2,Loophole4}.
The analysis above support  the main conclusion of
reference\citep{Kastner}: ``the delayed-choice quantum eraser neither
erases nor delays''. In particular, the successive measurement of
the $H$-polarization of the environment photon does not allow to
obtain the which-path information on the system photon. If, for instance,
detector D3 clicks it means that there is a fully costructive interference
between the two paths $a$  and $b$ at output 3 and a fully destructive
interference at output 4. These interferences can occur only if the system photon has passed through both the arms of the interferometer.

There is another possible scenario that has been suggested by a referee and that was not taken into account in this paper and in the previous ones. In this scenario the environment photon is detected after the system photon has passed through the PBS on the right, but before the system photon reaches the
subsequent beam-splitter on its way to detection. In such a case,  the interpretations of the quantum erasure appears evident here. However, we note that in this case too, there is no inversion of the natural  temporal order between the physical events related to the detections of the entangled photons. It is the first  detection of the environment photon that produces a quantum collapse of the entangled state  and, thus, determines the result of the next  detection of the system photon.

So far  we have analyzed
the special case of a delayed-choice quantum erasure for a photon-photon
entangled state but  analogous considerations could be applied
to any other entangled state of two sub-systems 1 and 2. Assume $|\psi_{12}\text{\textrangle}$
is a pure entangled state of two sub-systems 1 and 2 and $A$ and
$B$ are two observables for system 1 and 2, respectively. We indicate
by $|A_{i}\text{\textrangle}$ ($i$=1,$\infty$) and $|B_{j}\text{\textrangle}$  ($j$=1,$\infty$)  a complete
set of orthonormal eigenstates of the observables $A$ and $B$, respectively. Quantum
Mechanics predicts that the probability that the pure state $|\psi_{12}\text{\textrangle}$
is found in the state $|A_{K}B_{L}\text{\textrangle}=|A_{K}\text{\textrangle}|B_{L}\text{\textrangle}$
is 

\begin{equation}
p_{KL}=\left|\text{\textlangle}A_{K}|\text{\textlangle}B_{L}|\psi_{12}\text{\textrangle}\right|^{2},\label{eq:Smax-3-1-1-3-3-1}
\end{equation}
where state $|\psi_{12}\text{\textrangle}$ in the basis  $|A_{i}\text{\textrangle}|B_{j}\text{\textrangle}$ writes 
\begin{equation}
|\psi_{12}\text{\textrangle}= \sum_ {i=1}^{\infty}  \sum_ {j=1}^{\infty}\text{\textlangle}A_{i}|\text{\textlangle}B_{j}|\psi_{12}\text{\textrangle} |A_{i}\text{\textrangle}|B_{j}\text{\textrangle}.\label{eq:SmaxA}
\end{equation}
The calculation of the probability $p_{KL}$  in Eq.(\ref{eq:Smax-3-1-1-3-3-1})
could be  performed in two successive steps. Assume, for instance, that the observable $A$ of system 1 is first measured. The probability that  system 1   is found in the state  $|A_{K}\text{\textrangle}$ is:
\begin{equation}
p_{K}= \sum_ {j=1}^{\infty}\left|\text{\textlangle}A_{K}|\text{\textlangle}B_{j}|\psi_{12}\text{\textrangle}\right|^{2}\label{eq:SmaxC}
\end{equation}
According to Eq.(\ref{eq:SmaxA}), the normalized  state of system 2 after this  measurement is
\begin{equation}
|\psi_{2}\text{\textrangle}=  \frac{ \sum_ {j=1}^{\infty}\text{\textlangle}A_{K}|\text{\textlangle}B_{j}|\psi_{12}\text{\textrangle}|B_{j}\text{\textrangle}}{\sqrt{p_{K}}}.\label{eq:SmaxB}
\end{equation}
From Eq.(\ref{eq:SmaxB})  we infer that the conditional probability  $ p_{K|L}$ that system 2 if found in the state  $|B_{L}\text{\textrangle}$ after system 1 is found in the state  $|A_{K}\text{\textrangle}$ is
\begin{equation}
p_{K|L} =\frac{\left|\text{\textlangle}A_{K}|\text{\textlangle}B_{L}|\psi_{12}\text{\textrangle}\right|^{2}}{p_{K} } .\label{eq:SmaxD}
\end{equation}
Then, the joint probability $ p_{KL}$ to find both the states  $|A_{K}\text{\textrangle}$ and $|B_{L}\text{\textrangle}$  is   $p_{KL}= p_{K} \times p_{K|L}$ that coincides with the initial expression  of   $p_{KL}$  given  in  Eq.(\ref{eq:Smax-3-1-1-3-3-1}). The same conclusions are reached if we consider the alternative case where  observable $B$ is first measured on system 2. Then, the order of the measurements does not affect the value of the joint probability  $ p_{KL}$  but, according to the discussion above,  we think that the natural time ordered sequence
has to be physically preferred because it corresponds to the actual
order of the collapses of the entangled state.

\section{CONCLUSIONS}

In conclusion, we have shown  that the interpretation of the delayed-choice
experiments with entangled systems does not necessarily imply the
counterintuitive conclusion that the natural temporal order between
cause and effect has been reversed. Although the Quantum Mechanics
formalism in  Eq.(\ref{eq:Smax-3-1-1-3-3-1}) does not account for the temporal order of the measurements,
the delayed-choice results should be analyzed tacking into account
for the natural temporal order of the measurements. In particular,
in the case of the delayed-choice experiment shown in figure \ref{Fig 2},
it is the first detection of the system photon by detector D3 (or
D4) that determines the polarization state of the entangled environment
photon and  the correspondent successive detections by detectors D1 and
D2. Furthermore, no which-path information is allowed in this delayed case.  It has to be emphasized that the analysis above greatly
depends on the Quantum Mechanics assumption that the collapse of the
wave function occurs instantaneously everywhere in the space. In particular,
in the previous analysis it was  assumed that the environment
photon collapses instantaneously to the states $|E\text{\textrangle}_{e}$
 or $|E^{\bot}\text{\textrangle}_{e}$  when the system photon
is collected by detectors D3 and D4, respectively. Without the assumption
of an instantaneous collapse (or at the least of a sufficiently fast superluminal collapse),
the recent experimental results\citep{Kaiser,erasure} on the delayed-choice
quantum erasure performed in conditions of full space-like separation
between the investigated events (random choice and detections of the environment photon and
of the system photon) would not have been possible. The non-locality
of the quantum measurement process is in apparent contrast with the
Relativity theory and this  is just the more controversial aspect of
Quantum Mechanics that leads to some well known paradoxes and, in particular, to the EPR( Einstein, Podolsky and Rosen) paradox\citep{EPR}. Some physicists
are unsatisfied with the non-locality of Quantum Mechanics that is
in contrast with the predictions of any other previous theory as well
as the Maxwell electromagnetic theory and the Relativity theory. Alternative
local models based on local hidden variables have been proposed in
the past but the recent high accuracy  EPR experiments\citep{Loophole1,Loophole2,Loophole4} on the Bell-inequality\citep{Bell} have
definitely invalidated any hidden variables model also closing
the main  residual loopholes. In
more recent years, some physicists\citep{Eberhard_1989,Bohm_undivided_1993}
proposed alternative local models where  the correlations between entangled particles would be
established by superluminal signals propagating in a preferred
frame. The existence of a preferred frame where the superluminal signals
propagate isotropically in the space is needed to avoid the known
causal paradoxes\citep{Cocciaro_2013_ShutYourselfUp,Cocciaro3_DICE2015}.
In the limit case of an infinite velocity of the superluminal communications,
these local superluminal models lead to the same predictions of Quantum Mechanics and,
thus, no experiment can definitely invalidate them.
In this limit case, the choice between Quantum Mechanics
and superluminal models would only be a matter of taste. On the contrary,
it has been shown\citep{Salart_nature_2008} that there are  special experimental
configurations where discrepancies could be experimentally evidenced if
the velocity of the superluminal communications had a finite
value lower than a maximum detectable velocity that is imposed by  the  features of the experimental apparatus. Some experiments have been performed in recent years\citep{Salart_nature_2008,Cinesi_PhysRevLett2013,PhysRevA.97.052124} to evidence
these discrepancies but, so far, the predictions of Quantum Mechanics have been always confirmed.
As stated above, these negative results did not allow to  invalidate the local superluminal
models but  only allowed  to establish lower bounds for the possible
values of the superluminal velocities up to a few  million times the speed of  light.
\begin{acknowledgments}
I  acknowledge Bruno Cocciaro for helpful and stimulating discussions and for the critical reading of the manuscript and Massimo d'Elia  for the reading of a part of the manuscript.
\end{acknowledgments}

\bibliographystyle{apsrev4-1}
\bibliography{mybibPRA}

\end{document}